\documentclass[11pt,a4paper]{article}
\usepackage{amsmath,amssymb,amsfonts,a4wide,graphicx,bm,times}
\usepackage{mcite}

\usepackage{hyperref}
\hypersetup{
    colorlinks,
    citecolor=green,
    filecolor=green,
    linkcolor=blue,
    urlcolor=green
}

\makeatletter \let\old@startsection=\@startsection
\renewcommand{\@startsection}[6]
{\old@startsection{#1}{#2}{#3}{#4}{#5}{#6\mathversion{bold}}}
\makeatother

\makeatletter

\@addtoreset{equation}{section}
\makeatother

\newcommand{\superp}[2]{\genfrac{}{}{0pt}{}{#1}{#2}}

 \def\d{\delta}

 \def\Im{{\rm Im ~}}
 \def\p{\partial}
 
 \def\a{\alpha}
 \def\b{\beta}
 
 \def\d{\delta}
 \def\e{\varepsilon}

 \def\l{\lambda}

 \def\x{\xi}

 \def\s{\sigma}

 \def\G{\Gamma}

 \def\o{\omega }

\def\CC{{\mathcal{C}}}

\def\CF{{\mathcal{F}}}

\def\CN{{\mathcal{N}}}
\def\CO{{\mathcal{O}}}

\def\CS{{\mathcal{S}}}

\def\CY{{\mathcal{Y}}}

\def\Equskip{\!\!\!\!\!\!\!\!\!\!\!\!\!\!\!\!} 
\def\equskip{\!\!\!\!\!\!\!\!} 
\def\la{\left\langle}

\def\hf{\dfrac{1}{2}}

\def\CS{\mathcal{S}}

\def\ud{^{\uparrow\downarrow}}
\def\up{^\uparrow}
\def\down{^\downarrow}
\def\qf{\mathfrak{q}}
\def\Zv{\mathcal{Z}_\text{vect.}}

\def\Zf{\mathcal{Z}_\text{fund.}}

\def\Zinst{\mathcal{Z}_\text{inst.}}
\def\CYY{\CY_{\vec Y}}

\def\hCYY{\hat\CY_{\vec Y}}
\def\rag{\right\rangle_\text{gauge}}
\def\rahg{\right\rangle_{\widehat{\text{gauge}}}}
\def\hq{\hat q}
\def\hXi{\hat\Xi}
\def\hv{\hat v}
\def\hl{\hat l}
\def\ZZ{$\mathbb{Z}_2$}

\def\hG{\hat\Gamma}
\def\hZinst{\widehat\Zinst}
\def\hphi{\hat\phi}

\def\hg{\hat\gamma}

\begin{document}
\let\refOld\ref
\renewcommand{\ref}[1]{(\refOld{#1})}

\begin{titlepage}
\renewcommand{\thefootnote}{\fnsymbol{footnote}}
\vspace*{-2cm}
\begin{flushright}
KIAS-Q17054
\end{flushright}

\vspace*{1cm}

\begin{center}
{\huge {\bf  Quantum integrability of $\mathcal{N}=2$ 4d gauge theories}}

\vspace{10mm}
{\Large Jean-Emile Bourgine$^\dagger$$^\ast$, Davide Fioravanti$^\ast$}
\\[.6cm]
{\em {}$^\dagger$ Korea Institute for Advanced Study (KIAS)}\\
{\em Quantum Universe Center (QUC)}\\
{\em 85 Hoegiro, Dongdaemun-gu, Seoul, South Korea}\\
[.3cm]
{\em {}$^\ast$Sezione INFN di Bologna, Dipartimento di Fisica e Astronomia,
Universit\`a di Bologna} \\
{\em Via Irnerio 46, 40126 Bologna, Italy}
\\[.4cm]
\texttt{bourgine\,@\,kias.re.kr,\quad fioravanti\,@\,bo.infn.it}
\end{center}

\vspace{0.7cm}

\begin{abstract}
\noindent
We provide a description of the quantum integrable structure behind the Thermodynamic Bethe Ansatz (TBA)-like equation derived by Nekrasov and Shatashvili (NS) for $\mathcal{N}=2$ 4d Super Yang-Mills (SYM) theories. In this regime of the background, -- we shall show --, the instanton partition function is characterised by the solution of a TQ-equation. Exploiting a symmetry of the contour integrals expressing the partition function, we derive a 'dual' TQ-equation, sharing the same T-polynomial with the former. This fact allows us to evaluate to $1$ the quantum Wronskian of two dual solutions (for $Q$) and, then, to reproduce the NS TBA-like equation. The latter acquires interestingly the deep meaning of a known object in integrability theory, as its two second determinations give the usual non-linear integral equations (nlies) derived from the 'dual' Bethe Ansatz equations.

\vspace{0.5cm}
\end{abstract}

\vfill

\end{titlepage}
\vfil\eject

\setcounter{footnote}{0}

%
%
%

\section{Introduction}
The proposal \cite{Nekrasov2009} by Nekrasov and Shatashvili (NS) to associating quantum integrable systems to $\CN=2$ 4d Super Yang-Mills (SYM) theories in a particular regime of the Omega background has led numerous new insights in gauge theories. But also the possibility of applying the correspondence to solve some peculiar relativistic quantum integrable systems \cite{Grassi2014,Hatsuda2015,Codesido2015,Gu2015,Sciarappa2017}. The Omega background is known to depend on two parameters $\e_1$ and $\e_2$ that regularise the infinite volume of Euclidean space-time while preserving supersymmetry. Importantly, the ordinary background $\mathbb{R}^4$ is recovered in the limit $\e_1,\e_2\to0$ and yields the celebrated Seiberg-Witten (SW) theory \cite{Seiberg1994, Seiberg1994a}, whose BPS sector is described by a classical integrable system ({\it cf.} for instance  \cite{Marshakov1999} and references therein). Furthermore, Nekrasov and Shatashvili investigated the regime $\e_2\to0$ while $\e_1$ stays fixed and observed a quantisation of the SW integrable systems with $\e_1$ as Planck constant \cite{Nekrasov2009}.

In fact, it is important to distinguish two different integrable structures in the NS regime of Super Yang-Mills theories. The first one describes the moduli space of quantum vacua of the gauge theory. It is obtained by minimising with respect to the Higgs field  vacuum expectation values (vevs) the full prepotential: the latter, in fact, is formed by a perturbative (one loop) part and a non-perturbative part (sum on instantons). On the Coulomb branch, the Higgs field is given by the scalar field of the $\CN=2$ gauge vector multiplet, and the corresponding vevs are usually denoted by $a_l$ (the index $l$ running over the number of colors). In the SW limit, the minimisation reproduces their fundamental period relations \cite{Seiberg1994,Seiberg1994a}, linked to classical integrable systems \cite{Marshakov1999}. These are the systems quantized in the proposal of Nekrasov and Shatashvili \cite{Nekrasov2009,Nekrasov2009b} where the Coulomb branch vevs play the role of Bethe roots.

Yet, here we are interested in a second integrable structure, involving only the non-perturbative instanton contribution to the prepotential. In \cite{Nekrasov2009}, this contribution is written in terms of an integral equation \footnote{In the gauge theory context, the explicit derivation of this equation has been given in \cite{Meneghelli2013,Bourgine2014} (see also \cite{Bourgine2015,Bourgine2015a} for the sub-leading terms).}, of the same form as a Thermodynamic Bethe Ansatz (TBA) equation as initially obtained with temperature by Yang and Yang \cite{YY} or with finite size of a relativistic field theory (upon mirror transformation) by Al. B. Zamolodchikov \cite{Zamolodchikov1990a} \footnote{{\it Cfr.} also \cite{TBA1,TBA2,TBA3} for a general (non-relativistic) extension.}.

In this paper, we aim at providing a microscopic description of the degrees of freedom giving rise to this equation. In this description, the Coulomb branch vevs $a_l$ will be considered as free parameters, playing the role of inhomogeneities. On the other hand, Bethe roots will appear as the heights of the instanton stacks, more precisely as the heights of the columns of the Young diagrams describing these stacks.

One of the interesting aspects of this second type of integrable structure is its intricate connection with the further affinization of the algebras behind standard quantum integrable models \cite{Kanno2013,Nekrasov2013,Bourgine2014a,Bourgine2015c}. In the present case of 4d $\CN=2$ gauge theories, the relevant algebraic structure has been constructed by Vasserot and Schiffmann in \cite{Schiffmann2012}. It is built upon the Double Degenerate Affine Hecke algebra, and is called Spherical Hecke central (SHc) algebra. It really corresponds to the affinization of the Yangian algebra of non-relativistic/isotropic integrable systems like the XXX spin chain \cite{Tsymbaliuk2014}.\footnote{A quantum deformation of this structure arise in the five dimensional uplifts of these theories, namely 5d $\CN=1$ quiver Super Yang-Mills theories. The corresponding algebra is called quantum toroidal algebra of $\mathfrak{gl}(1)$, or Ding-Iohara-Miki algebra \cite{Ding1997,Miki2007}. Its realization in the BPS sector of the gauge theories has been studied in \cite{Mironov2016,Awata2016,Awata2016a,Bourgine2016,Bourgine2017b}.}

The results presented in this paper were largely inspired by the work of Poghossian et al. \cite{Poghossian2010,Fucito2011,Fucito2012} in which the NS regime is described by a TQ-like equation (in fact the $hv$ equation below\footnote{Actually, $h$ is an entire function here, as it should be, but $v$ is not: this is the reason why we called this a TQ-like equation.}). Until now, the relation between this TQ-like equation and the TBA-like equation of Nekrasov and Shatashvili remained unclear. This is one of the main questions that motivated the present work. The two key ingredients of the answer are the presence of a genuine (with $Q$ entire) TQ equation and of a reflection symmetry leading to its 'dual' TQ-equation. The corresponding $hv$ equations are obtained by sending the cut-off to infinity in the thermodynamical limit upon suitable regularisation (which introduces the pole structure in $v$). Then, the TBA-like equation is recovered by taking the quantum Wronskian relation of these two equations. Eventually, its second determinations \cite{Destri1997} in the upper and lower half-planes, respectively, do coincide with the counting functions of the non-linear integral equations (nlies) derived from the two dual sets of Bethe Ansatz equations ({\it cf.} for instance \cite{Klumper1991,Destri1994,Fioravanti1996}). In conclusion, the NS non-linear integral equation is given a physical meaning very different from a thermodynamic set-up. Of course, the Bethe equations can be associated to the two dual TQ-equations.

The paper is organised as follows. Starting from the instanton partition function of the gauge theory localised in the Omega-background, we observe an invariance under the deformation of integration contours. This invariance provides alternative expressions for the instanton partition function, and the qq-character. The later is defined as a correlator of the gauge theory, it is a polynomial of degree equal to the number of colors. These quantities will be defined in section two. Then, the NS limit $\e_2\to0$ will be performed in section three using a minimization procedure, namely a saddle point approximation of the sum over instanton configurations. In this limit, the qq-character reproduces the Baxter T-polynomial of the TQ-equation. The dual TQ-equation is obtained from the reflection symmetry mentioned previously. In this procedure, we will introduce some cut-offs interpreted as the number of roots in each Bethe string. Then, we will perform the thermodynamical limit, sending these cut-offs to infinity. In this limit, the Baxter TQ-equations are replaced by some equations, here called '$hv$' equations, involving non-polynomial quantities. Finally, we will show that the quantum Wronskian relation of these equations reproduces the NS TBA-like equation. In addition, the section three also contains a brief discussion on non-linear integral equations and the second determinations of the TBA-like equation. The latter is definitively associated to nlies equivalent to some Bethe Ansatz equations rather than to a thermodynamic procedure on Bethe Ansatz equations, a fully different physical meaning.

\section{Instanton partition function and qq-characters}\label{sec_Zinst}
\subsection{Instanton partition function}
In this paper, we will provide a description of the NS regime for 4d $\CN=2$ Super Yang-Mills theories with a gauge group $U(N_c)$, and $N_f$ matter fields in the fundamental (or anti-fundamental) representation. For technical reasons, our analysis is restricted to the cases $N_f<2N_c$ for which the theories are asymptotically free. In the NS regime, the theories depend on the Omega background parameter $\e_1$ while the second parameter $\e_2$ is sent to zero. The gauge theory is taken on the Coulomb branch, and the vacuum expectation values (vevs) of the scalar field in the gauge multiplet are denoted $a_l$ with the index $l$ running over the number of colors $N_c$. Masses of the matter fields are denoted $m_f$ with $f$ running over the number of flavors $N_f$. These parameters $a_l$ and $m_f$ are encoded in the roots of the gauge and matter polynomials defined respectively as
\begin{equation}\label{def_AMQ}
A(z)=\prod_{l=1}^{N_c}(z-a_l),\quad M(z)=\prod_{f=1}^{N_f}(z-m_f),\quad Q(z)=\dfrac{M(z)}{A(z)A(z+\e_+)}.
\end{equation}
We have also introduced the rational potential $Q(z)$ that plays an essential role in the integral formulation of the instanton partition function.

The instanton partition function is obtained after localization as an instanton expansion in terms of the (renormalized) exponentiated gauge coupling $\qf$ \cite{Nekrasov2003}. \footnote{The sign of the gauge coupling parameter $\qf$ has been reversed with respect to the convention used in \cite{Bourgine2015c}, but it coincides with the notation employed in \cite{Bourgine2014}.} It takes the form of a sum over the $n$-instantons sectors of coupled integrals,
\begin{equation}\label{def_Zinst}
\Zinst=\sum_{n=0}^\infty\dfrac{\qf^n}{n!}\left(\dfrac{\e_+}{\e_1\e_2}\right)^n\oint_\G{\prod_{\superp{i,j=1}{i<j}}^nK(\phi_{ij})\ \prod_{i=1}^n Q(\phi_i)\dfrac{d\phi_i}{2i\pi}},
\end{equation} 
where we denoted $\e_+=\e_1+\e_2$. We have also employed the shortcut notation $\phi_{ij}=\phi_i-\phi_j$. The integration contour $\G$ goes on $(-\infty,+\infty)$ and circles the upper half-plane while avoiding the infinity. In this expression, the background parameters $\e_1$ and $\e_2$ have a positive imaginary part, and the Coulomb branch vevs are assumed to be real with a small positive imaginary part, as in $a_l+i0$, so that poles at $\phi_i=a_l$ are taken inside the contour, but not those at $\phi_i=a_l-\e_+$. The kernel $K$ that couples the integration variables depends only on the background parameters, it is a rational function that can be written in a factorized form,
\begin{equation}\label{def_KS}
K(z)=\dfrac{z^2(z^2-\e_+^2)}{(z^2-\e_1^2)(z^2-\e_2^2)}=S(z)^{-1}S(-z)^{-1},\quad S(z)=\dfrac{(z+\e_1)(z+\e_2)}{z(z+\e_+)}.
\end{equation} 
The scattering factor $S(z)$ plays a key role in the quantum algebra behind the qq-characters \cite{Bourgine2015c}, it obeys the property $S(-z)=S(z-\e_+)$. Finally, the variables $\phi_i$ are assumed to be real in the kernels $K(\phi_{ij})$, so that poles $\phi_i=\phi_j+\e_\a$ ($\a=1,2$) are inside the contours, but not the poles at $\phi_i=\phi_j-\e_\a$.  

The contour integrals in the expression \ref{def_Zinst} can be evaluated using Cauchy theorem as a sum of residues. The poles in the instanton sector $n$ are in one-to-one correspondence with the box configurations of $N_c$ Young diagrams $\vec Y=(Y^{(1)},\cdots, Y^{(N_c)})$ having a total number of boxes $|\vec Y|=n$. Precisely, the poles are located at the positions $\phi_x=a_l+(i-1)\e_1+(j-1)\e_2$ associated to the boxes $x=(l,i,j)$ of coordinates $(i,j)$ in the $l$th Young diagram $Y^{(l)}$. The corresponding residue can be factorized into contributions of gauge and matter multiplets, and the instanton partition function written formally as
\begin{equation}\label{Zinst_Ydiag}
\Zinst=\sum_{\vec Y}(-\qf)^{|\vec Y|}\Zv(\vec a,\vec Y)\Zf(\vec m,\vec Y),
\end{equation} 
where $\vec Y$ are $N_c$-tuples Young diagrams, and $\vec a$, $\vec m$ vectors of dimension $N_c$ and $N_f$ respectively, encoding the dependence in Coulomb branch vevs and fundamental multiplets masses.

The vector and fundamental contributions to the residues obey a set of recursion equations that describes the insertion or removal of an instanton in the stack described by the configuration $\vec Y$,
\begin{equation}\label{Ward}
\dfrac{\Zv(\vec a,\vec Y+x)}{\Zv(\vec a,\vec Y)}=-\dfrac1{\e_1\e_2}\dfrac{\prod_{y\in R(\vec Y)}\phi_{xy}(\phi_{xy}-\e_+)}{\prod_{\superp{y\in A(\vec Y)}{y\neq x}}\phi_{xy}(\phi_{xy}+\e_+)},\quad\dfrac{\Zf(\vec m,\vec Y+x)}{\Zf(\vec m,\vec Y)}=M(\phi_x).
\end{equation} 
Here, $\vec Y+x$ (resp $\vec Y-x$) denotes the Young diagram $\vec Y$ with the box $x$ added (removed). Accordingly, the sets $A(\vec Y)$ and $R(\vec Y)$ contains the boxes that can be added to or removed from the Young diagrams $\vec Y$. These equation have been referred to as \textit{discrete Ward identities}, they first appeared in the study \cite{Kanno2013} of the representation of the Spherical Hecke central algebra \cite{Schiffmann2012} on instanton partition functions. They were written in the form \ref{Ward} in \cite{Bourgine2014a,Bourgine2015c} where a short derivation can also be found.

The original expressions for the contributions $\Zf$ and $\Zv$ are fairly complicated. Yet, it is possible to write simpler expressions by solving the discrete Ward identities. The fundamental contribution is simply the product of matter polynomials evaluated at the location of the residues. The vector contribution is slightly more complicated,
\begin{equation}\label{def_Zv}
\Zf(\vec m,\vec Y)=\prod_{x\in\vec Y}M(\phi_x),\quad \Zv(\vec a,\vec Y)=\prod_{x\in\vec Y}\left[A(\phi_x+\e_+)\CYY(\phi_x)\right]^{-1},
\end{equation} 
it involves the function $\CYY(z)$ defined as
\begin{equation}\label{def_CYY}
\CYY(z)=\dfrac{\prod_{x\in A(\vec Y)}(z-\phi_x)}{\prod_{x\in R(\vec Y)}(z-\e_+-\phi_x)}.
\end{equation}
In the following, we will not need these expressions, but only work at the level of the Ward identities \ref{Ward}.

\subsection{Definition of the qq-character}
In addition to the partition function, another quantity will be helpful to characterize the NS regime of the theory. Actually, this quantity has been introduced by Nekrasov in full generality \cite{Nekrasov2015,Nekrasov2016} (see also \cite{Nekrasov2013,Nekrasov2017}) under the name of \textit{qq-character}. It provides a generalization of the quantum group q-characters defined in \cite{Knight1995} for Yangians, and in \cite{Frenkel1998} for quantum affine algebras. This quantity possesses the crucial property of being a polynomial in the spectral variable $z$, which derives by implementing the non-perturbative Schwinger-Dyson equations (as loop equations on a suitable resolvent). In fact, the qq-character is defined as a trace over the instanton configurations,
\begin{equation}\label{def_chi}
\chi(z)=\la\CYY(z+\e_+)-\qf \dfrac{M(z)}{\CYY(z)}\rag\quad \text{with}\quad \la\CO_{\vec Y}\rag=\dfrac1{\Zinst}\sum_{\vec Y}(-\qf)^{|\vec Y|}\Zv(\vec a,\vec Y)\Zf(\vec m,\vec Y) \CO_{\vec Y} 
\end{equation} 
where the function $\CYY(z)$ defined in \ref{def_CYY} depends on the detailed content of the Young diagrams. 

Moreover, it enjoys another expression in terms of contour integrals,
\begin{equation}\label{chi_int}
\chi(z)=\dfrac{A(z+\e_+)}{\Zinst}\sum_{n=0}^\infty\dfrac{\qf^n}{n!}\left(\dfrac{\e_+}{\e_1\e_2}\right)^n\oint_{\G-C_{z+\e_+}}{\prod_{\superp{i,j=1}{i<j}}^nK(\phi_{ij})\ \prod_{i=1}^n S(z-\phi_i)Q(\phi_i)\dfrac{d\phi_i}{2i\pi}}.
\end{equation} 
This expression of the qq-character reproduces the initial definition \ref{def_chi} after an evaluation of the contour integral by Cauchy theorem. The prescription for the integration is to consider $z$ with a small positive imaginary part, so that it is inside the integration contour $\G$, but to exclude the point $z+\e_+$ by circulating it (clock-wise) with the small circle $C_{z+\e_+}$. In this way, we take only the extra pole at $\phi_i=z$ in the integration contour, not the one at $\phi_i=z+\e_+$. Evaluation of the residues brings two different types of terms, corresponding to the sum of the two terms in the RHS in the first equation of \ref{def_chi}. The first term is obtained when no variable $\phi_i$ takes the extra pole value $\phi_i=z$, so that the residues contribute as
\begin{equation}
\dfrac{A(z+\e_+)}{\Zinst}\sum_{\vec Y}(-\qf)^{|\vec Y|}\Zv(\vec a,\vec Y)\Zf(\vec m,\vec Y)\prod_{x\in \vec Y}S(z-\phi_x)=\la\CYY(z+\e_+)\rag.
\end{equation} 
When $n>1$, the second possibility is that one of the variables $\phi_i$ takes the value $\phi_i=z$, bringing the residue
\begin{equation}
-\dfrac{\e_1\e_2}{\e_+}Q(z)\oint_\G{\prod_{\superp{i,j=1}{i<j}}^{n-1}K(\phi_{ij})\ \prod_{i=1}^{n-1}\dfrac{Q(\phi_i)}{S(z-\phi_i-\e_+)}\dfrac{d\phi_i}{2i\pi}}.
\end{equation} 
New poles appear at $\phi_i=z-\e_\a$ for $\a=1,2$, but they are outside the contour $\G$ since $z$ is assumed to be slightly above the real axis. Hence, the remaining pole configurations are in one-to-one correspondence with the $|\vec Y|=n-1$ boxes in the $N_c$-tuple of Young diagrams. Summing over the instanton sectors gives the second term in the RHS of the first equation in \ref{def_chi}:
\begin{align}
\begin{split}
&\dfrac{A(z+\e_+)}{\Zinst}\sum_{n=1}^\infty\dfrac{\qf^n}{n!}\left(\dfrac{\e_+}{\e_1\e_2}\right)^n\cdot n\cdot\left( -\dfrac{\e_1\e_2}{\e_+}\right)Q(z)\sum_{\superp{\vec Y}{|\vec Y|=n-1}}\left(-\dfrac{\e_1\e_2}{\e_+}\right)^{|\vec Y|}\Zv(\vec a,\vec Y)\Zf(\vec m,\vec Y)\prod_{x\in\vec Y}\dfrac1{S(z-\phi_x-\e_+)}=\\
=&-\qf\dfrac{A(z+\e_+)}{\Zinst}Q(z)\sum_{\vec Y}(-\qf)^{|\vec Y|}\Zv(\vec a,\vec Y)\Zf(\vec m,\vec Y)\prod_{x\in\vec Y}\dfrac1{S(z-\phi_x-\e_+)}=-\qf\la\dfrac{M(z)}{\CYY(z)}\rag.
\end{split}
\end{align}

\subsection{Reflection symmetry}
The instanton partition function $\Zinst$ turns out to be invariant under a \ZZ-symmetry that can be seen from a deformation of contours in the integral expression \ref{def_Zinst}. Explicitly, the contour $\G$ is deformed on the Riemann sphere into the contribution of the point at infinity and the contour $\hG$ that circles the lower half-plane (excluding infinity). Defining the following quantity\footnote{The extra minus sign $(-1)^n$ comes from the inversion of the orientation of the contours, so that $\hG$ is oriented anti-clock-wise.}
\begin{equation}\label{def_Zinst_hat}
\hZinst=\sum_{n=0}^\infty\dfrac{\qf^n}{n!}\left(-\dfrac{\e_+}{\e_1\e_2}\right)^n\oint_{\hG}{\prod_{\superp{i,j=1}{i<j}}^nK(\phi_{ij})\ \prod_{i=1}^n Q(\phi_i)\dfrac{d\phi_i}{2i\pi}},
\end{equation} 
we have $\Zinst=\hZinst$ by contour deformation, except in the case $N_f=2N_c-1$ where poles at $\phi_i=\infty$ appear. The corresponding residues can be easily worked out, they factorize in front:
\begin{equation}
\Zinst=e^{\qf\e_+/(\e_1\e_2)}\hZinst\quad \text{for} \ N_f=2N_c-1.
\end{equation} 
The quantity $\hZinst$ can also be evaluated by Cauchy theorem, with poles still in bijection with the boxes of the $N_c$-tuple Young diagrams $\vec Y$, but now located at the dual positions $\hphi_x=a_l-i\e_1-j\e_2$ for each $x=(l,i,j)\in\vec Y$. As a result, we find the weak coupling expansion
\begin{align}
\begin{split}
\hZinst=&\sum_{\vec Y}(-\qf)^{|\vec Y|}\widehat{\Zv(\vec a,\vec Y)}\widehat{\Zf(\vec m,\vec Y)},\\
\text{with:}&\\
&\widehat{\Zf(\vec m,\vec Y)}=\prod_{x\in\vec Y}M(\hphi_x),\quad \widehat{\Zv(\vec a,\vec Y)}=\prod_{x\in\vec Y}\left[A(\hphi_x)\hCYY(\hphi_x)\right]^{-1}.
\end{split}
\end{align}
Here we have denoted $\hCYY(z)$ the \ZZ-dual of the function $\CYY(z)$:
\begin{equation}
\hCYY(z)=\dfrac{\prod_{x\in A(\vec Y)}(z-\hat\phi_x)}{\prod_{x\in R(\vec Y)}(z+\e_+-\hat\phi_x)}.
\end{equation}

It turns out that the dual quantities $\hZinst$, $\widehat{\Zf(\vec m,\vec Y)}$ and $\widehat{\Zv(\vec a,\vec Y)}$ can be obtained from the original ones using a simple \ZZ-symmetry. This symmetry consists in a sign flip of the Omega-background parameters $\e_1,\e_2$, followed by a shift of the Coulomb branch vevs $a_l\to a_l-\e_+$. As a result, the poles located at $z=\phi_x$ in the upper half-plane are mapped to the dual position $z=\hphi_x$ in the lower half-plane. Under this symmetry, the gauge polynomials $A(z)$ and $A(z+\e_+)$ are exchanged, and the argument of the scattering factor is shifted/reversed: $S(z)\to S(z-\e_+)=S(-z)$, effectively sending $\widehat{\Zv(\vec a,\vec Y)}$ to $\Zv(\vec a,\vec Y)$ and vice versa. Note that in the case of a $U(1)$ gauge theory, it is possible to show that $\widehat{\Zv(\vec a,\vec Y)}=\Zv(\vec a,\vec Y)$, but this is no longer true for a higher number of colors.

The deformation of the contour can also be applied to the integral expression \ref{chi_int} of the qq-character. The new contour does not contain the pole at $\phi_i=z$, but instead it includes the extra pole at $\phi_i=z+\e_+$ that had been subtracted from $\G$ previously:
\begin{equation}
\chi(z)=\dfrac{A(z+\e_+)}{\hZinst}\sum_{n=0}^\infty\dfrac{\qf^n}{n!}\left(\dfrac{\e_+}{\e_1\e_2}\right)^n\oint_{\hG+ C_{z+\e_+}}{\prod_{\superp{i,j=1}{i<j}}^nK(\phi_{ij})\ \prod_{i=1}^n S(z-\phi_i)Q(\phi_i)\dfrac{d\phi_i}{2i\pi}}.
\end{equation} 
In the case $N_f=2N_c-1$, the contribution of poles at infinity is the same as in the case of the instanton partition function, and the extra factors cancel. The evaluation using the Cauchy theorem follows the same lines as what was done in the previous section, and
\begin{equation}
\chi(z)=\la\hCYY(z)-\qf \dfrac{M(z+\e_+)}{\hCYY(z+\e_+)}\rahg,\quad \text{with}\quad \la \CO_{\vec Y} \rahg=\dfrac1{\hZinst}\sum_{\vec Y}(-\qf)^{|\vec Y|}\widehat\Zv(\vec a,\vec Y)\widehat\Zf(\vec m,\vec Y)\ \CO_{\vec Y}
\end{equation} 
where the average is taken with respect to configurations of poles $\vec Y$ located in the lower half-plane. Shifting the spectral parameter $z\to z-\e_+$, it is possible to rewrite this second expression of the qq-character in a form that makes the \ZZ-symmetry more explicit:
\begin{equation}\label{dual_qq}
\chi(z-\e_+)=\la\hCYY(z-\e_+)-\qf \dfrac{M(z)}{\hCYY(z)}\rahg.
\end{equation} 
Thus, the \ZZ-symmetry described previously simply exchanges $\chi(z)$ and $\chi(z-\e_+)$.

In the limit $\b=1$ of the background where $\e_+=0$, the instanton positions are given by the simplified expression $\phi_x=a_l+\e_1(i-j)$ and the \ZZ-symmetry corresponds to exchange Young diagrams with their transposed. In the Seiberg-Witten limit $\e_1,\e_2\to0$, the \ZZ-symmetry is trivially realized, so that $\phi_x$ and $\hphi_x$ condense to form the same branch cut.


Before ending this discussion on the \ZZ-symmetry, we would like to mention that it is closely related to another reflection symmetry that consists in the following transformation:
\begin{equation}
a_l\to \e_+-a_l,\quad \qf\to (-1)^{N_f}\qf,\quad m_f\to -m_f.
\end{equation} 
In the $SU(2)$ case, this symmetry coincides with the reflection symmetry of Liouville theory that exchanges the operators of charge $\a$ and $Q-\a$. This second symmetry exchanges $A(z)\leftrightarrow(-1)^{N_c}A(-z+\e_+)$ but the scattering factor $S(z)$ remains invariant. It acts on the poles location by sending $\phi_x\to-\hphi_x$ and $\hphi_x\to-\phi_x$, which leads to the following transformation for the instanton contributions:
\begin{equation}
\Zf(\vec m,\vec Y)\to(-1)^{N_f|\vec Y|}\widehat{\Zf(\vec m,\vec Y)},\quad \Zv(\vec a,\vec Y)\to\widehat{\Zv(\vec a,\vec Y)},
\end{equation} 
so that the instanton partition function $\Zinst$ is again sent to $\hZinst$ (it is invariant if $N_f<2N_c-1$). The functions $\CYY(z)$ and $(-1)^{N_c}\hCYY(-z)$ are exchanged so that the qq-character $\chi(z)$ transforms into $(-1)^{N_c}\chi(-z-\e_+)$. Hence we observe that the action of this second \ZZ-symmetry is very similar to the previous one, up to extra minus sign factors and a reflection of the spectral variable $z$. Although the physical interpretation of the second \ZZ-symmetry is more straightforward, it is in fact more convenient to work with the first one 
because of the absence of these extra signs.

\section{The structures of integrability in the Nekrasov-Shatashvili limit}
\subsection{Minimization procedure}
It is now well-known how to perform the Nekrasov-Shatashvili limit $\e_2\to0$ on the expression \ref{Zinst_Ydiag} of the partition function as a sum over Young diagrams \cite{Nekrasov2013,Bourgine2014a,Bourgine2015c}. However, for a matter of completeness, we will recall it here. The justification of this procedure has been given in \cite{Nekrasov2013}. The main idea behind it is that the sum over Young diagrams becomes dominated by a single ($N_c$-tuple) diagram $\vec Y^\ast$, in a sort of discrete saddle point.  This diagram contains infinitely many boxes, arranged in columns of infinite height, but keeping a fixed number $n_l$ of columns in each diagram $Y^{\ast(l)}$. The critical diagram must obey the saddle point equation expressing that the small variation corresponding to add (or remove) a box has no cost at first order in $\e_2$:
\begin{equation}
(-\qf)^{|\vec Y^\ast+x|}\Zv(\vec a,\vec Y^\ast+x)\Zf(\vec m,\vec Y^\ast+x)=(-\qf)^{|\vec Y^\ast|}\Zv(\vec a,\vec Y^\ast)\Zf(\vec m,\vec Y^\ast) +O(\e_2).
\end{equation} 
Obviously, this equation can only be satisfied by infinitely large diagrams. Using the discrete Ward identity \ref{Ward}, this condition takes the form
\begin{equation}\label{saddle}
\dfrac{\qf}{\e_1\e_2}M(\phi_x)\dfrac{\prod_{y\in R(\vec Y^\ast)}\phi_{xy}(\phi_{xy}-\e_+)}{\prod_{\superp{y\in A(\vec Y^\ast)}{y\neq x}}\phi_{xy}(\phi_{xy}+\e_+)}=1.
\end{equation} 
Next, we assume that a box $x$ can be added to every column of $\vec Y^\ast$. The associated pole locations are $\phi_x=a_l+(i-1)\e_1+\l_i^{(l)}\e_2$ where $\l_i^{(l)}$ is the length of the column to which $x$ is added, here the $i$th column of the $l$th diagram. We will denote these quantities $u_r$ with the double index $r=(l,i)$. They will later play the role of Bethe roots. Under this assumption, below each box in $A(\vec Y)$ is a box in $R(\vec Y)$ that can be removed, with approximately the same pole location (up to a negligible $\e_2$ shift). This is true, except for a subset of boxes in $A(\vec Y)$ that are located at the extreme right of the diagram, i.e. in the nearest empty column of each diagram $Y^{\ast(l)}$. The pole location corresponding to these boxes will be denoted $\xi_l=a_l+n_l\e_2$. Then, the product in the saddle point equation \ref{saddle} can be replaced by
\begin{equation}\label{Bethe}
\dfrac{\qf M(u_r)}{\Xi(u_r)\Xi(u_r+\e_1)}\prod_{s=1}^M\dfrac{u_r-u_s-\e_1}{u_r-u_s+\e_1}=1,\quad \Xi(z)=\prod_{l=1}^{N_c}(z-\xi_l).
\end{equation} 
We recognize here a set of Bethe equations, with a total number $M=\sum_ln_l$ of Bethe roots. These are well-defined Bethe equations and they can be solved exactly in $\qf$ for a small number of Bethe roots.

Although the previous system of Bethe equation is well-defined for a finite number of roots $M$, and finite cut-offs $\xi_l$, one should keep in mind the perturbative aspect of the problem in the application to the NS regime of the gauge theory. By construction, in the relevant solution, the Bethe roots are arranged in $N_c$ strings, each of length $n_l$ and spaced of $\e_1$. Hence, in the limit $\qf\to0$, the solution becomes trivial as $u_r$ tends to $u_r^{(0)}=a_l+(i-1)\e_1$ for $r=(l,i)$.\footnote{Effectively as if $\vec Y^\ast$ were empty, but one should be careful with the order of limits.} Formally, the prepotential in the NS limit $\CF_\text{NS}$ can be reconstructed from the solution of the Bethe equations using the formula
\begin{equation}\label{prepotential}
\qf\dfrac{\p\CF_\text{NS}}{\p\qf}=\lim_{\e_2\to0}\e_1\e_2|\vec Y|=\lim_{n_l\to\infty}\e_1\sum_{r=1}^M(u_r-u_r^{(0)}),\quad\text{with}\quad \CF_\text{NS}=\lim_{\e_2\to0}\e_1\e_2\log\Zinst
\end{equation} 
Employing an argument due to Poghossian \cite{Poghossian2010}, it is shown in appendix \refOld{AppA} that the subleading correction to $u_r$ is of order $O(\qf^i)$ for $r=(l,i)$. It implies that solutions with $n_l$ or $n_l+1$ roots in the $l$th string will only differ at the order $O(\qf^{n_l+1})$. Thus, in order to recover the exact prepotential, it is necessary to consider the thermodynamical limit where the number $n_l$ of roots in each string (i.e. the number of columns in each Young diagram composing $\vec Y^\ast$) is infinite. More precisely, in order to compute the prepotential at the order $O(\qf^N)$, it is necessary and sufficient to consider $N$ roots in each string, in which case $M=N_cN$. This fact is illustrated in appendix \refOld{AppA} in the simple case of a pure $U(1)$ gauge theory. In a sense, the integer variables $n_l$ play the role of cut-offs that must be sent to infinity at the end of our computation. The thermodynamic limit will be discussed in more details in the next subsection.

A Baxter TQ-equation can be written for the set of Bethe equations \ref{Bethe}\footnote{Taking the pure gauge case $N_f=0$, it is possible to send $\xi_l$ to infinity while keeping the number of roots fixed (although it doesn't make much sense in our model). The divergent factors $\xi_l$ can be absorbed by the introduction of an essential singularity at the infinity for the Q-polynomial: $t(z)\to \prod_l\xi_l t(z)$ and $q(z)\to \prod_l\xi_l^{-z/\e_1} q(z)$. Then, this equation reduces to the Baxter TQ-equation relevant for the Toda system with twisted periodic boundary conditions,
\begin{equation}
t(z)q(z)=q(z+\e_1)-\qf q(z-\e_1).
\end{equation}
We will not take this approach here. The connection between the Toda chain and the NS NLIE has been studied in \cite{Kozlowski2010}.}
\begin{equation}\label{TQ}
t(z)q(z)=\Xi(z)\Xi(z+\e_1)q(z+\e_1)-\qf M(z)q(z-\e_1),\quad q(z)=\prod_{r=1}^M(z-u_r).
\end{equation} 
The derivation is done in appendix \refOld{AppA}. This equation is well-defined for a finite number of roots: both $t(z)$ and $q(z)$ are polynomials, and the expansion in $z$ provides enough constraints to determine them from the knowledge of $M(z)$ and $\Xi(z)$. Actually, the TQ-equation \ref{TQ} can also be derived from the definition \ref{def_chi} of the qq-character by taking its NS limit \cite{Bourgine2015c}. In this limit, the trace of well-behaved operators is dominated by a single term, $\la\CO_{\vec Y}\rag\simeq \CO_{\vec Y^\ast}$, where $\vec Y^\ast$ is the solution to the saddle point equation described previously. In particular, the operator $\CYY(z)$ defined in \ref{def_CYY}, and its inverse, produce ratios of the Baxter Q-polynomial,
\begin{equation}
\la\CYY(z)\rag\simeq\dfrac{q(z)\Xi(z)}{q(z-\e_1)},\quad \la\dfrac1{\CYY(z)}\rag\simeq\dfrac{q(z-\e_1)}{q(z)\Xi(z)}.
\end{equation} 
Denoting $\bar\chi(z)$ the polynomial of degree $N_c$ obtained by taking the NS limit of $\chi(z)$, we find from \ref{def_chi},
\begin{equation}
\bar\chi(z)=\dfrac{q(z+\e_1)\Xi(z+\e_1)}{q(z)}-\qf M(z)\dfrac{q(z-\e_1)}{q(z)\Xi(z)},
\end{equation} 
leading to identify $t(z)=\bar\chi(z)\Xi(z)$. This identification is in agreement with the $\qf$-perturbative study of the TQ-equation \ref{TQ} done in appendix \refOld{AppA} where it is shown that $t(z)=h(z)\Xi(z)+O(\qf^{1+\min_ln_l})$. Terms of order $O(\qf^{1+\min_ln_l})$ are negligible in the thermodynamical limit where all $n_l$ are sent to infinity, and we will later identify $\bar\chi(z)=h(z)$.

\subsection{Thermodynamical limit}
In order to regularize the infinite product of Bethe roots defining the Q-polynomial, we introduce the function $q_0(z)$ that corresponds to the limit $\qf\to0$ of $q(z)$,
\begin{equation}\label{def_q0}
q_0(z)=\prod_{r=1}^M(z-u_r^{(0)})=\prod_{l=1}^{N_c}\prod_{i=1}^{n_l}(z-a_l-(i-1)\e_1).
\end{equation} 
It obeys the important property
\begin{equation}\label{prop_q0}
\dfrac{q_0(z-\e_1)}{q_0(z)}=\dfrac{\Xi(z)}{A(z)}.
\end{equation}
This polynomial $q_0(z)$ can be used to define the ratio $\bar v(z)=q(z)/q_0(z)$ that tends to $v(z)$ in the thermodynamic limit. As a consequence of the TQ equation \ref{TQ} and the property \ref{prop_q0}, it obeys the following difference equation:
\begin{equation}
\dfrac{t(z)}{\Xi(z)}\bar v(z)=A(z+\e_1)\bar v(z+\e_1)-\qf \dfrac{M(z)}{A(z)} \bar v(z-\e_1).
\end{equation}
It is shown in the appendix \refOld{AppA} that the T-polynomial in the $n_l\to\infty$ limit behaves as
\begin{equation}
\dfrac{t(z)}{\Xi(z)}\to h(z),
\end{equation} 
where $h(z)$ is a monic polynomial of degree $N_c$ that coincides with the limit of the qq-character. In this limit, the functions $h$ and $v$ obey a difference equation that is reminiscent of a TQ-equation,\footnote{
A similar equation has been obtained in a seemingly different manner in \cite{Poghossian2010,Fucito2011}. In order to make a precise connection, one can introduce the function $\o(z)=v(z-\e_1)/(v(z)A(z))$ and show that it obeys 
\begin{equation}
\qf M(z-\e_1)\o(z)\o(z-\e_1)+h(z-\e_1)\o(z)=1.
\end{equation}}
\begin{equation}
h(z)A(z)v(z)=A(z)A(z+\e_1)v(z+\e_1)-\qf M(z)v(z-\e_1).
\end{equation} 
However, it is noted that $v(z)$ is no longer an entire function here, it exhibits some poles at $z=a_l+(i-1)\e_1$ with $l=1\cdots N_c$ and $i\in\mathbb{Z}^{>0}$. Introducing the rational potential $Q(z)$ defined in \ref{def_AMQ} (with $\e_+\to\e_1$), we can write the ``$hv$'' equation:
\begin{equation}\label{hv}
\dfrac{h(z)}{A(z+\e_1)}v(z)=v(z+\e_1)-\qf Q(z)v(z-\e_1).
\end{equation} 

\subsection{Dual TQ and $hv$ equations}
The same saddle point procedure can be applied to the dual expression of the qq-character given in \ref{dual_qq}. In the NS limit, the vev of the function $\hCYY(z)$ becomes
\begin{align}
\begin{split}
&\la\hCYY(z)\rahg\simeq \dfrac{\hq(z)\hXi(z)}{\hq(z+\e_1)},\quad \la\dfrac1{\hCYY(z)}\rahg\simeq \dfrac{\hq(z+\e_1)}{\hq(z)\hXi(z)},\\
&\text{with:}\quad \hq(z)=\prod_{l=1}^{N_c}\prod_{i=1}^{n_l}(z-a_l+i\e_1+\e_2\hat\l_i^{(l)}),\quad \hXi(z)=\prod_{l=1}^{N_c}(z-a_l+(\hat n_l+1)\e_1)
\end{split}
\end{align}
Note however that the $N_c$-tuple Young diagram extremizing the summations of the gauge and $\widehat{\text{gauge}}$ brackets will be different. Thus, the Bethe roots $u_r$ and the dual ones $\hat u_{r=(l,i)}=a_l-i\e_1-\e_2\hat\l_i^{(l)}$ will be unrelated. From the expression \ref{dual_qq} of the qq-character, we deduce a new (dual) tq-equation, related to the previous one under the \ZZ-symmetry described above,
\begin{equation}\label{dual_TQ}
\hat t(z-\e_1)\hq(z)=\hXi(z)\hXi(z-\e_1)\hq(z-\e_1)-\qf M(z)\hq(z+\e_1).
\end{equation}
with $\hat t(z)=\bar\chi(z)\hXi(z+\e_1)$. This is again a well-defined TQ-equation, where all the quantities are polynomials. Interestingly, it takes a form similar to the original TQ-equation \ref{TQ}, with the sign of the pseudo-period $\e_1$ reversed, and the polynomial $\Xi(z)$ replaced by $\hXi(z)$. In fact, the minimization procedure would provide a set of Bethe equations in which the same replacement occurs.

The procedure to perform the thermodynamical limit is the same as in the case of the original TQ-equation. We introduce the function $\hq_0$ obtained in the $\qf\to0$ limit of $\hq$,
\begin{equation}
\hq_0(z)=\prod_{l=1}^{N_c}\prod_{i=1}^{\hat n_l}(z-a_l+i\e_1),\quad \dfrac{\hq_0(z+\e_1)}{\hq_0(z)}=\dfrac{\hXi(z)}{A(z+\e_1)},
\end{equation} 
and write \ref{dual_TQ} for the ratio $\hq/\hq_0\to\hv$. As $n_l$ is sent to infinity, we obtain the dual $hv$ equation
\begin{equation}
\dfrac{h(z-\e_1)}{A(z)}\hv(z)=\hv(z-\e_1)-\qf Q(z) \hv(z+\e_1).
\end{equation} 
The resemblance with the original $hv$-equation strikes even more: the ratio $h(z)/A(z+\e_1)$ has been shifted by $-\e_1$ while the sign of the pseudo-period has been reversed in the arguments of the function $v(z)$.

\subsection{Quantum Wronskian and Non-Linear Integral Equation}
The relation between the NS TBA-like equation and a quantum Wronskian has already been mentioned in \cite{Kozlowski2010}. However, it appears that the underlying quantum systems are different. Furthermore, the presence of a reflection symmetry bringing a second TQ-equation is essential in our derivation, and this ingredient seems to be missing in \cite{Kozlowski2010}.

In order to establish the invariance of the quantum Wronskian, we start from the two $hv$-equations derived previously,
\begin{align}
\begin{split}
&\dfrac{h(z)}{A(z+\e_1)}v(z)=v(z+\e_1)-\qf Q(z)v(z-\e_1),\\
&\dfrac{h(z-\e_1)}{A(z)}\hv(z)=\hv(z-\e_1)-\qf Q(z) \hv(z+\e_1).
\end{split}
\end{align}
These two equations can be solved for $h(z)$, and we find the two equalities
\begin{equation}
h(z)=A(z+\e_1)\dfrac{v(z+\e_1)}{v(z)}-\qf\dfrac{M(z)}{A(z)}\dfrac{v(z-\e_1)}{v(z)}=A(z+\e_1)\dfrac{\hv(z)}{\hv(z+\e_1)}-\qf \dfrac{M(z+\e_1)}{A(z+2\e_1)}\dfrac{\hv(z+2\e_1)}{\hv(z+\e_1)}.
\end{equation} 
After multiplication of both sides by $v(z)\hv(z+\e_1)/A(z+\e_1)$, the second equality can be written in the form
\begin{equation}
v(z+\e_1)\hv(z+\e_1)-\qf Q(z)v(z-\e_1)\hv(z+\e_1)=v(z)\hv(z)-\qf Q(z+\e_1)v(z)\hv(z+2\e_1).
\end{equation} 
Defining the quantum Wronskian to be
\begin{equation}
W(z)=v(z)\hv(z)+\qf Q(z)v(z-\e_1)\hv(z+\e_1),
\end{equation} 
we have thus shown that $W(z)=W(z+\e_1)$. By construction, $v(z)$ and $\hv(z)$ are perturbative series in $\qf$, and meromorphic in $z$ at each order of $\qf$. It implies that $W(z)$ is a constant that can be determined from the fact that $v(z)\sim 1$ and $\hv(z)\sim1$ while $Q(z)\sim z^{N_f-2N_c}$ at infinity. Hence, we have $W(z)=1$, i.e.
\begin{equation}\label{qWronskian}
v(z)\hv(z)+\qf Q(z)v(z-\e_1)\hv(z+\e_1)=1.
\end{equation} 

In order to make contact with the NLIE derived by Nekrasov and Shatashvili \cite{Nekrasov2009} (see also \cite{Meneghelli2013,Bourgine2014}), we need to introduce the kernel function
\begin{equation}
G(z)=\dfrac1{z+\e_1}-\dfrac1{z-\e_1}.
\end{equation}
and the integration contour $\G$ which is the one involved in the integral expression of the Nekrasov partition function \cite{Nekrasov2003}. This contour circles the upper half-plane, including the real axis but excluding the point at infinity. In the evaluation of contour integrals, the Omega-background parameters $\e_1$ and $\e_2$ are considered with a positive imaginary part. In general, the result takes the form of a rational function of these parameters and it can be easily analytically continued. In addition, the Coulomb branch vevs are assumed to possess a small imaginary part so that $a_l$ belong to the contour $\G$ while $a_l-\e_1$ do not. As a result, the Bethe roots $u_r$ and the string positions $u_r^{(0)}$ are inside the integration contour $\G$, while the dual roots $\hat u_r$ and positions $\hat u_r^{(0)}$ belong to the deformed contour $\hat\G$ circling the lower half-plane and such that $\G\cup\hat\G=\{\infty\}$. This trick is used in appendix \refOld{AppB} to show perturbatively in $\qf$ that
\begin{equation}
\oint_\G G(z-w)\log(v(w)\hv(w))\dfrac{dw}{2i\pi}=-\log(v(z-\e_1)\hv(z+\e_1)),
\end{equation}
where the spectral parameter $z$ has a small positive imaginary part as above. Introducing the function 
\begin{equation}\label{def_eps}
e^{-\e(z)}=1-v(z)\hv(z),
\end{equation} 
the quantum Wronskian equation can be written in the form of a non-linear integral equation,
\begin{equation}\label{NS_NLIE}
e^{-\e(z)}=\qf Q(z)\exp\left(-\oint_\G G(z-w)\log(1-e^{-\e(w)})\right),
\end{equation} 
which is exactly the one proposed by Nekrasov and Shatashvili. By construction, its solution is $\mathbb{Z}_2$-symmetric and because of the analogy with Yang-Yang-Zamolodchikov TBA equation (except for the integration contour), it has been called \textit{pseudo-energy}.

\subsection{The NLIE procedure and the second determination}
A procedure based on the Cauchy theorem \cite{Destri1994, Fioravanti1996} can be applied to the TQ-equation \ref{TQ}, defining the counting function $\bar\eta(z)$ and the associated resolvent $r(z)$
\begin{equation}\label{def_counting}
e^{2i\pi\bar\eta(z)}=\dfrac{\qf M(z)}{\Xi(z)\Xi(z+\e_1)}\dfrac{q(z-\e_1)}{q(z+\e_1)},\quad r(z)=\p_z\log(1-e^{2i\pi\bar\eta(z)}).
\end{equation} 
Using the TQ-equation, the resolvent is expressed in the form
\begin{equation}
r(z)=\p_z\log\left(\dfrac{t(z)q(z)}{\Xi(z)\Xi(z+\e_1)q(z+\e_1)}\right).
\end{equation}
We deduce from this expression that the resolvent has poles with residues $+1$ at $z=u_r$ (Bethe roots) and at the zeros of $t(z)$ (holes) that were denoted $e_l\down,e_l\up$ in appendix \refOld{AppA}. In addition, poles with residues $-1$ are present at $z=u_r-\e_1$, $z=\xi_l$ and $z=\xi_l-\e_1$. Taking the integral with a contour $\CC(u_r,u_r-\e_1)$ that the surrounds the Bethe roots and their shifted value, we can write
\begin{equation}
\sum_{r=1}^{M}\left[\log(z-u_r-\e_1)-\log(z-u_r+\e_1)\right]=\oint_{\mathcal{C}(u_r,u_r-\e_1)}\Equskip\equskip\log\left((z-w-\e_1)(z-w)\right) r(w)\dfrac{dw}{2i\pi}.
\end{equation}
Here we will assume $\Im z<-\Im\e_1$ in order to avoid the branch cut singularities of the logarithm. Taking the log of the definition \ref{def_counting} and replacing the RHS of the previous equation, we find
\begin{equation}
2i\pi\bar\eta(z)=\log\left(\dfrac{\qf M(z)}{\Xi(z)\Xi(z+\e_1)}\right)+\oint_{\mathcal{C}(u_r,u_r-\e_1)}\Equskip\equskip\log\left((z-w-\e_1)(z-w)\right)r(w)\dfrac{dw}{2i\pi}.
\end{equation}

The next step is to take the thermodynamic limit in order to write an integral equation for the quantity $2i\pi\eta(z)$ obtained as the limit of $2i\pi\bar\eta(z)$:
\begin{equation}\label{eta_v}
e^{2i\pi\bar\eta(z)}\to e^{2i\pi\eta(z)}=\qf Q(z)\dfrac{v(z-\e_1)}{v(z+\e_1)},
\end{equation} 
The equality in the RHS has been obtained after the introduction of $q_0(z)$ in \ref{def_counting} to turn $\Xi(z)$ into the gauge polynomial $A(z)$. Then, the ratios $q(z)/q_0(z)$ have been replaced by $\bar v(z)$ that tends to $v(z)$ in the thermodynamic limit. In this limit, the holes $e_l\up=\xi_l+O(\qf^{n_l+1})$ will tend to coincide with the singularities of $r(z)$ at $z=\xi_l$. Likewise, the holes $e_l\down$ will become very close to the Bethe roots $u_{l,1}-\e_1$ that are inside the integration contour. In order to avoid the integration contour being pinched between poles, it is necessary to include the holes within the integration contour. At the same time, it will be more convenient to also add the singularities at $z=\xi_l$ and $z=\xi_l+\e_1$ in order to modify the driving term, and write
\begin{equation}
2i\pi\bar\eta(z)=\log\left(\dfrac{\qf M(z)\Xi(z)\Xi(z-\e_1)}{t(z)t(z-\e_1)}\right)+\oint_{\G-\e_1}\log\left[(z-w-\e_1)(z-w)\right]r(w)\dfrac{dw}{2i\pi},
\end{equation}
where the new integration contour, denoted formally $\G-\e_1$, corresponds to the contour $\G$ shifted downward by $-\Im\e_1$ so that it encompass all the singularities of the resolvent $r(z)$. It is now possible to take the thermodynamic limit both in the driving term and in the contour integral. Using an integration by part, the resulting integral equation for the thermodynamical counting function $2i\pi\eta(z)$ reads
\begin{equation}\label{NLIE_I}
2i\pi\eta(z)=\log\left(\dfrac{\qf M(z)}{h(z)h(z-\e_1)}\right)+\oint_{\G-\e_1} k(z-w)\log\left(1-e^{2i\pi\eta(w)}\right)\dfrac{dw}{2i\pi},\quad k(z)=\dfrac1z+\dfrac1{z-\e_1},
\end{equation} 
where the spectral parameter is assumed to satisfy $\Im z<-\Im\e_1$, i.e. it lies outside the contour $\G-\e_1$.

The same procedure can be applied to the dual TQ-equation, and we find that the thermodynamical counting function defined as
\begin{equation}\label{heta_hv}
e^{2i\pi\hat\eta(z)}=\qf Q(z)\dfrac{\hv(z+\e_1)}{\hv(z-\e_1)},
\end{equation} 
obeys the integral equation
\begin{equation}\label{NLIE_II}
2i\pi\hat\eta(z)=\log\left(\dfrac{\qf M(z)}{h(z)h(z+\e_1)}\right)-\oint_{\hat\G+\e_1} \hat k(z-w)\log\left(1-e^{2i\pi\hat\eta(w)}\right)\dfrac{dw}{2i\pi},\quad \hat k(z)=\dfrac1z+\dfrac1{z+\e_1},
\end{equation} 
where $\hat\G$ surrounds the lower half-plane, and $\hat\G+\e_1$ is shifted upward by $\Im\e_1$.

The NLIE established by Nekrasov and Shatashvili is defined for the range $-\Im\e_1<\Im z<\Im\e_1$ of the spectral parameter. It is possible to consider the second determination of the function $\e(z)$ in the half-planes $\Im z<-\Im\e_1$ and $\Im z>\Im\e_1$, denoted respectively $\e_-(z)$ and $\e_+(z)$. They are obtained by considering the RHS of the integral equation \ref{NS_NLIE} for a spectral parameter $z$ outside of the principal domain $|\Im z|<\Im\e_1$. Accordingly, the integral kernel of NLIE $G(z-w)$ picks up an extra pole, either at $w=z+\e_1$ or at $w=z-\e_1$:
\begin{equation}\label{2nd_det}
e^{-\e_-(z)}=\dfrac{e^{-\e(z)}}{1-e^{-\e(z+\e_1)}},\quad e^{-\e_+(z)}=\dfrac{e^{-\e(z)}}{1-e^{-\e(z-\e_1)}}.
\end{equation}
In the RHS of these two formulas, the pseudo-energy $\e(z)$ is analytically continued to the lower or upper half planes $\Im z<-\Im\e_1$ and $\Im z>\Im\e_1$ respectively. The analytic continuation is provided using the definition \ref{def_eps} expressed in terms of the function $v(z)$ and $\hv(z)$ satisfying the quantum Wronskian equation \ref{qWronskian}. When expressed in terms of the function $v(z)$ and $\hv(z)$, the RHS of the two formulas in \ref{2nd_det} reproduce the formulas obtained in \ref{eta_v} and \ref{heta_hv}, and expressing $\eta(z)$ and $\hat\eta(z)$ in terms of $v(z)$ and $\hv(z)$ respectively. It shows that the two thermodynamical counting functions $\eta(z)$ and $\hat\eta(z)$ are equal to the two second determinations of the pseudo-energy (modulo $2i\pi$), namely $2i\pi\eta(z)=-\e_-(z)[2i\pi]$ and $2i\pi\hat\eta(z)=-\e_+(z)[2i\pi]$.

We would like to end this section with a practical remark. After deformation of the integration contour for the NLIE \ref{NLIE_I} the counting function $\eta(z)$ is seen to satisfy the following functional equation,
\begin{equation}
e^{2i\pi\eta(z)}=\qf M(z)\dfrac{1-e^{2i\pi\eta(z)}}{h(z)}\dfrac{1-e^{2i\pi\eta(z-\e_1)}}{h(z-\e_1)},
\end{equation} 
where $h(z)$ is the limit of the qq-character $\chi(z)$. Thus, once the qq-character is known,\footnote{For $N_c=2$ $N_f=0$ $a_1=a_2=a$, the expression of the qq-character is known at all orders in $\qf$ \cite{Bourgine2015c}:
\begin{equation}
\chi(z)=(z+\e_+)^2-a^2+\e_1\e_2\qf\p_{\qf}\log\Zinst.
\end{equation}} these equations can be solved perturbatively in the gauge coupling parameter $\qf$,\footnote{This expansion is taken with $h(z)$ finite although it does also depend on $\qf$.}
\begin{equation}
e^{2i\pi\eta(z)}=\sum_{n=1}^\infty\qf^nH_n(z),\quad H_1(z)=\dfrac{M(z)}{h(z)h(z-\e_1)},\quad H_{n+1}(z)=H_1(z)\sum_{k=0}^n H_k(z)H_{n-k}(z-\e_1),
\end{equation}
and $H_0(z)=-1$. Inverting the relation between $\eta(z)$ and the second determination $\e_-(z)$, this method provides one of the most efficient way to compute the pseudo energy $\e(z)$:
\begin{equation}
e^{-\e(z)}=\sum_{n=1}^\infty E_n(z),\quad E_0(z)=-1,\quad E_{n+1}(z)=-\sum_{k=0}^n H_{k+1}(z)E_{n-k}(z-\e_1).
\end{equation} 

\section{Summary and discussion}
In this paper, we have presented the integrable structure behind the TBA-like equation by Nekrasov and Shatashvili \cite{Nekrasov2009}, which arises when evaluating the $\e_2\to0$ limit of the sum over instanton configurations. In fact, this regime is a natural quantisation/regularisation of SW theory \cite{Seiberg1994, Seiberg1994a} as $\e_1$ is still finite and is characterised by a set of Bethe equations with a polynomial Baxter TQ equation. Exploiting the reflection symmetry realised as a deformation of the integration contours, a dual TQ equation has been written. In the thermodynamical limit, the two TQ equations produce a dual pair of difference equations respectively, the $hv$-equations, albeit the $v$ functions are no longer entire. Yet, their quantum Wronskian reproduces the TBA-like equation above \cite{Nekrasov2009} for a sort of pseudo-energy. In addition, we have investigated the non-linear integral equations obeyed by the counting functions for the Bethe roots. Very interestingly, these counting functions coincide with the two second determinations of the NS pseudo-energy.

The reflection symmetry introduced here is very close to the reflection symmetry of Liouville vertex operators. But it is possible to give it another algebraic interpretation. Lifting up the theory to a five dimensional background compactified on $S^1$, the corresponding $\CN=1$ gauge theory is covariant under the action of the Ding-Iohara-Miki algebra \cite{Ding1997,Miki2007} in the instanton sector \cite{Mironov2016,Awata2016,Awata2016a,Bourgine2016,Bourgine2017b}. This q-deformation of the SHc algebra simplifies the identification of the action of symmetries by lifting some degeneracies. In this context, the \ZZ-symmetry presented here is expected to coincide with the reflection $\s_H$ defined in \cite{Bourgine2017d}, and that acts on the Drinfeld currents as follows:\footnote{This expectation is partially based on the property $\CYY(-z)=-\hCYY(z-\e_+)$ for $N_c=1$ and $a_1=0$.}
\begin{equation}
x^\pm(z)\to x^\pm(1/z),\quad \psi^\pm(z)\to\psi^\mp(1/z),\quad \hg\to\hg.
\end{equation}
This symmetry maps the algebra DIM$_{q_1,q_2}$ to DIM$_{q_1^{-1},q_2^{-1}}$, the inversion of the parameters $q_1,q_2$ corresponding to a sign flip of the parameters $\e_1,\e_2$ in the degenerate case. Similarly, the inversion of the spectral parameter becomes a change of sign $z\to -z$ as the radius of $S^1$ is set to zero. In the correspondence with $(p,q)$-web diagrams of IIB string theory, the symmetry $\s_H$ acts as a reflection of the diagram with the axis in the direction associated to D5-branes. This fact seems coherent with the interpretation as a reflection in Liouville theory. Yet, some further investigations are required to properly establish all these claims. We hope to address this issue in a near future.

In a companion paper \cite{Bourgine2017c}, we investigated the double deformation of the Seiberg-Witten relations: at finite $\e_1$ and $\e_2$, we managed to identify the symmetry exchanging the two 'sheets' of the Seiberg-Witten curve. This symmetry is different from the one described here, and the solutions $v(z)$ and $\hat v(z)$ of the two dual $hv$-equations reproduce the same solution of the SW curve equation in the limit $\e_1\to0$. In fact, the difference between these two reflection symmetries is more easily understood in the five dimensional uplift theory. In this case, the exchange of the two SW sheets has been identified with the action of another reflection symmetry of the DIM generators, denoted by $\s_V$ in \cite{Bourgine2017d}.\footnote{Miki's automorphism $\CS$ \cite{Miki2007} relates the two reflections as $\s_V=\CS^2\s_H$. It is expected to be related to S-duality, and satisfies $\CS^4=1$.} In the $(p,q)$-brane language, $\s_V$ may be interpreted as a reflection with axis in the direction associated to NS5-branes.

In this paper, we have confined our discussion to the TBA-like and the non-linear integral equations, second determinations of the former. Still, it should be also possible to relate the prepotential in the limit $\e_2\to0$ to the Yang-Yang functionals associated to our systems of Bethe roots. Importantly, the prepotential obeys some $\hbar$-deformed Seiberg-Witten relations (with $\hbar=\e_1$) \cite{Bourgine2012,Bourgine2012a}, which, in their turn, shall be obtainable as NS limit of the 'doubly-quantised' qqSW relations ($\e_2\neq 0$) found in \cite{Bourgine2017c} in the full omega background. Therefore (and with crucial importance), the r\^ole of the Seiberg-Witten differential upon quantisation seems to be played by the counting function $\eta(z)$: this very intriguing point is currently under investigation.

Finally, if we have found a system of Bethe equations underlying the TBA-like equation, yet a physical model is still missing. The question of finding an Hamiltonian (and a more complete transfer-matrix) describing this system is still open. In fact, this quantum system could be obtained using the R-matrix found in \cite{Smirnov2013}, after taking the limit $\e_2\to0$. We hope to come back to this point soon.

\section*{Acknowledgments}

We would like to thank D. Bombardelli for his interest at the early stage of this project. We also thank F. Morales, V. Pasquier and D. Volin for stimulating discussions (especially on the quantum Wronskian). JEB thanks IPhT CEA-Saclay for warm hospitality. This project was partially supported by the grants: (I.N.F.N. IS) GAST (which supported JEB by an I.N.F.N. post-doctoral fellowship), UniTo-SanPaolo Nr TO-Call3-2012-0088, the ESF Network HoloGrav (09-RNP-092 (PESC)), MPNS--COST Action MP1210 and the EC Network Gatis.

\appendix

\section{Analysis of the Bethe equations}\label{AppA}
\subsection{Order of the corrections to the Bethe roots}
To show that the deviation of the Bethe roots from the string solution is of the form $u_{(l,i)}=u_{(l,i)}^{(0)}+O(\qf^i)$, we will work by induction on $k$ and assume the hypothesis $\d_{l,i}=u_{(l,i)}-u_{(l,i)}^{(0)}=O(\qf^i)$ for $i\leq k$ and $\d_{l,i}=o(\qf^k)$ for $i>k$. For this purpose, we also need to write the Bethe equations in the form
\begin{equation}\label{Bethe_app}
\Xi(u_{(l,i)})\Xi(u_{(l,i)}+\e_1)q(u_{(l,i)}+\e_1)=\qf M(u_{(l,i)}) q(u_{(l,i)}-\e_1).
\end{equation}
where $q(z)$ is the Q-polynomial defined in \ref{TQ}. We first need to show that our hypothesis is true for $k=1$. By definition, $\d_{l,i}=o(1)$ so that for $i>1$ the RHS of the Bethe equations is
\begin{equation}\label{Bethe_rhs}
\qf M(u_{(l,i)}) q(u_{(l,i)}-\e_1)=\qf M(u_{(l,i)})(\d_{l,i}-\d_{l,i-1})\prod_{s\neq (l,i-1)}(u_{(l,i)}-u_s-\e_1)=o(\qf),
\end{equation}
since $\d_{l,i}-\d_{l,i-1}$ is of order $o(1)$ and the remaining terms are of order one. Then, we examine the LHS. Taking $i=n_l$, we have
\begin{equation}\label{prop_xi}
\Xi(u_{(l,n_l)}+\e_1)=\d_{l,n_l}\prod_{l'\neq l}(u_{(l,n_l)}+\e_1-\x_{l'}),
\end{equation}
which implies, expanding in $\qf$,
\begin{equation}
\Xi(u_{(l,n_l)})\Xi(u_{(l,n_l)}+\e_1)q(u_{(l,n_l)}+\e_1)=\d_{l,n_l}\Xi(u_{(l,n_l)}^{(0)})\prod_{l'\neq l}(u_{(l,n_l)}^{(0)}+\e_1-\x_{l'})q_0(u_{l,n_l}^{(0)}+\e_1)+O(\qf).
\end{equation}
Thus the Bethe equations imply $\d_{l,n_l}=o(\qf)$. Then, we take $l$ and $i$ such that $1\leq i<n_l$, and consider the expansion of
\begin{equation}\label{prop_qs}
q(u_{(l,i)}+\e_1)=(\d_{l,i}-\d_{l,i+1})\prod_{s\neq (l,i+1)}(u_{(l,i)}-u_s+\e_1)=(\d_{l,i}-\d_{l,i+1})\prod_{s\neq (l,i+1)}(u_{(l,i)}^{(0)}-u_s^{(0)}+\e_1)+O(\qf)
\end{equation}
Since $\Xi(u_{(l,i)})\Xi(u_{(l,i)}+\e_1)$ remains finite, by the equality \ref{Bethe_app}, and \ref{Bethe_rhs} that gives the order of the RHS, we deduce that $\d_{l,i}=\d_{l,i+1}+o(\qf)$. By recursion, since $\d_{l,n_l}=o(\qf)$, we have $\d_{l,i}=o(\qf)$ for all $l,i$ with $i>1$. For $i=1$, the RHS of the Bethe equation is of order $O(\qf)$ since the product $Mq$ remains finite, while the LHS is proportional to $\d_{l,i}-\d_{l,i+1}$ according to \ref{prop_qs}. It implies that $\d_{l,1}=O(\qf)$. Thus, the induction hypothesis is true for $k=1$.

Assuming the hypothesis true at rank $k$, the induction follows the same steps. Using the decomposition of the product \ref{Bethe_rhs}, it is shown that when $i>k+1$, the RHS of the Bethe equations is of order $o(\qf^{k+1})$ and it is of order $O(\qf^{k+1})$ when $i=k+1$:
\begin{equation}
\qf M(u_{(l,i)})q(u_{(l,i)}-\e_1)=\qf(\d_{l,i}-\d_{l,i-1})M(u_{(l,i)}^{(0)})\left(\prod_{s\neq (l,i-1)}(u_{(l,i)}^{(0)}-u_s^{(0)}-\e_1)+O(\qf)\right).
\end{equation} 
Taking $i=n_l$, due to \ref{prop_xi} the LHS of the Bethe equations is equal to $\d_{l,n_l}$ times a function of order one which is non-vanishing. It implies that $\d_{l,n_l}=o(\qf^{k+1})$. Then, we consider $k+1<i<n_l$ and use the property \ref{prop_qs} to show that the LHS of the Bethe equation is equal to $(\d_{l,i}-\d_{l,i+1})$ times a non-vanishing function of order one. This shows that $\d_{l,i}=\d_{l,i+1}+o(\qf^{k+1})$ and by induction $\d_{l,i}=o(\qf^{k+1})$ for all $l$ and $i>k+1$. Finally, we take $i=k+1$ and due to the same property \ref{prop_qs} the LHS of the Bethe equation is again equal to $(\d_{l,k+1}-\d_{l,k+2})$ times a non-vanishing function of order one. But now the RHS is of order $O(\qf^{k+1})$, implying $\d_{l,k+1}=\d_{l,k+2}+O(\qf^{k+1})=O(\qf^{k+1})$. This finishes the proof of the hypothesis at rank $k+1$ and the induction.

\subsection{Explicit solution in the case $N=1$, $N_f=0$}
In the case of a pure $U(1)$ gauge theory, the NS prepotential has the simple form $\CF_{NS}=\qf$. The Bethe equations can be solved explicitly for a small number of roots. For two roots, we find after a bit of algebra,
\begin{equation}
u_1=a+\dfrac{3}{2}\e_1+\dfrac{2\qf\e_1}{\a}-\hf\sqrt{\a+\e_1^2},\quad u_2=a+\dfrac{3}{2}\e_1+\dfrac{2\qf\e_1}{\a}+\hf\sqrt{\a+\e_1^2},
\end{equation}
with $\a(\qf)$ solving the following cubic equation,
\begin{equation}
\a^3+4\qf\a^2+8\qf\e_1^2\a+16\e_1^2\qf^2=0.
\end{equation}
It can be solved order by order in $\qf$, thus providing a series for $\a(\qf)$ which leads to
\begin{align}
\begin{split}
&u_1(\qf)=a+\dfrac{\qf}{\e_1}+\dfrac{\qf^2}{2\e_1^3}+\dfrac{\qf^3}{2\e_1^5}+\dfrac{3\qf^4}{4\e_1^7}+\dfrac{11\qf^5}{8\e_1^9}+\dfrac{85\qf^6}{32\e_1^{11}}+O(\qf^7),\\
&u_2(\qf)=a+\e_1-\dfrac{\qf^2}{2\e_1^3}-\dfrac{3\qf^3}{4\e_1^5}-\dfrac{\qf^4}{\e_1^7}-\dfrac{23\qf^5}{16\e_1^9}-\dfrac{79\qf^6}{32\e_1^{11}}+O(\qf^7).
\end{split}
\end{align}
Taking the sum of $\e_1(u_r-u_r^{(0)})$ as in \ref{prepotential}, we recover the expansion of the exact prepotential up to the order $O(\qf^2)$. 

The solution of the Bethe equations involving three roots can be obtained perturbatively in $\qf$,
\begin{align}
\begin{split}
&u_1(\qf)=a+\dfrac{\qf}{\e_1}+\dfrac{\qf^2}{2\e_1^3}+\dfrac{7\qf^3}{12\e_1^5}+O(\qf^4),\\
&u_2(\qf)=a+\e_1-\dfrac{\qf^2}{2\e_1^3}-\dfrac{2\qf^3}{3\e_1^5}+O(\qf^4),\\
&u_3(\qf)=a+2\e_1+\dfrac{\qf^3}{12\e_1^5}+O(\qf^4).
\end{split}
\end{align}
We observe that the expansion of $u_1(\qf)$ and $u_2(\qf)$ coincide with the two-roots solution up to the order $O(\qf^2)$, and differ at the order $O(\qf^3)$. And now, the sum of $\e_1(u_r-u_r^{(0)})$ reproduces the expansion of the exact prepotential up to the order $O(\qf^3)$.

\subsection{Derivation of the TQ-equation}
The Bethe equations \ref{Bethe} can be rewritten in the following form,
\begin{equation}
\Xi(u_r)\Xi(u_r+\e_1)\prod_{s\neq r}\dfrac{u_r-u_s+\e_1}{u_r-u_s}=-\qf M(u_r)\prod_{s\neq r}\dfrac{u_r-u_s-\e_1}{u_r-u_s}.
\end{equation}
Consider the decomposition of the following functions over their poles at $z=u_r$ and $z=\infty$,
\begin{align}
\begin{split}
&\Xi(z)\Xi(z+\e_1)\prod_r\dfrac{z-u_r+\e_1}{z-u_r}=\e_1\sum_r\dfrac{\Xi(u_r)\Xi(u_r+\e_1)}{z-u_r}\prod_{s\neq r}\dfrac{u_r-u_s+\e_1}{u_r-u_s}+\left[\Xi(z)\Xi(z+\e_1)\dfrac{q(z+\e_1)}{q(z)}\right]_+,\\
&\qf M(z)\prod_r\dfrac{z-u_r-\e_1}{z-u_r}=-\qf\e_1\sum_r\dfrac{M(u_r)}{z-u_r}\prod_{s\neq r}\dfrac{u_r-u_s-\e_1}{u_r-u_s}+\left[\qf M(z)\dfrac{q(z-\e_1)}{q(z)}\right]_+,
\end{split}
\end{align}
where $q(z)$ denotes Baxter's Q-polynomial and the subscript $+$ the positive powers of $z$ in the expansion at infinity. We observe that the two functions have the same poles, and the same residues at $z=u_r$. They only differ by their singularities at $z=\infty$, which is a polynomial of degree $2N_c$ that we denote $t(z)$. Thus, we have obtained the TQ-equation \ref{TQ}, with the T-polynomial given by
\begin{equation}\label{expr_t}
t(z)=\left[\Xi(z)\Xi(z+\e_1)\dfrac{q(z+\e_1)}{q(z)}\right]_+-\left[\qf M(z)\dfrac{q(z-\e_1)}{q(z)}\right]_+.
\end{equation}

\subsection{Perturbative analysis of the T-polynomial}
The T-polynomials is a monic polynomial of degree $2N_c$, and we denote the roots $e_l\ud$ with $l=1\cdots N_c$. In the limit $\qf\to0$, $q(z)$ tends to $q_0(z)$ and due to the property \ref{prop_q0}, $t(z)$ tends to the product $\Xi(z)A(z+\e_1)$. This provides the position of the holes at first order in $\qf$,
\begin{equation}
e_l\up=\xi_l+O(\qf),\quad e_l\down=a_l-\e_1+O(\qf).
\end{equation}
These holes may be interpreted as the companion roots of the strings corresponding to $u_{(l,i)}$ with $l$ fixed. Actually, it is possible to show that $\xi_l$ is a zero of $t(z)$ with a much better approximation, that is $t(\xi_l)=O(\qf^{n_l+1})$. This is due to the fact that $\xi_l$ is an exact zero of $\Xi(z)$, and since $u_{(l,n_l)}=\xi_l-\e_1+\d_{l,n_l}$ and $\d_{l,n_l}$ is of order $O(\qf^{n_l})$ we have also
\begin{equation}
\dfrac{q(\xi_l-\e_1)}{q(\xi_l)}=\dfrac{\d_{l,n_l}}{\d_{l,n_l}-\e_1}\prod_{s\neq (l,n_l)}\dfrac{\xi_l-u_s-\e_1}{\xi_l-u_s}=O(\qf^{n_l}).
\end{equation}
Thus, the RHS of \ref{expr_t} for $z=\xi_l$ is indeed of order $O(\qf^{n_l+1})$. Setting $n=\min_ln_l$, we define $h(z)$ as the monic polynomial with roots at $z=e_l\down$, so that $t(z)=\Xi(z)h(z)+O(\qf^{n+1})$.

\section{Analytic properties of the $v$-functions and contour integration}\label{AppB}
We will work perturbatively in $\qf$. At first order, $v(z)=1+O(\qf)$ and the logarithm can be expanded formally as
\begin{equation}
\log(v(z))=\sum_{n=1}^\infty\qf^nl_n(z).
\end{equation} 
By construction, each function $l_n(z)$ has poles only at the values $z=a_l+(i-1)\e_1$ for $i\in\mathbb{Z}^{>0}$, so that they all lie in the contour of integration $\G$. Consider the integral 
\begin{equation}
\oint_\G G(z-w)\log(v(w))\dfrac{dw}{2i\pi}=\sum_{n=1}^\infty\qf^n\left[\oint_\G \dfrac{l_n(w)}{z-w+\e_1}\dfrac{dw}{2i\pi}-\oint_\G \dfrac{l_n(w)}{z-w-\e_1}\dfrac{dw}{2i\pi}\right].
\end{equation}
We can deform the integration contour on the sphere. At $z=\infty$, $v(z)=1+O(1/z)$, taking the logarithm and expanding, we deduce that $l_n(z)=O(1/z)$ and there is no pole at infinity. The integrals can be written with a contour $\hat\G$ that surrounds the lower-half plane, avoiding the real axis and the point at infinity,
\begin{equation}
\oint_\G G(z-w)\log(v(w))\dfrac{dw}{2i\pi}=-\sum_{n=1}^\infty\qf^n\left[\oint_{\hat\G} \dfrac{l_n(w)}{z-w+\e_1}\dfrac{dw}{2i\pi}-\oint_{\hat\G} \dfrac{l_n(w)}{z-w-\e_1}\dfrac{dw}{2i\pi}\right].
\end{equation} 
The variable $z$ is assumed to lie slightly above the real axis, and the first integral has no pole inside the new integration contour. The second integral has only one pole, located at $w=z-\e_1$, and with the residue $-l_n(z-\e_1)$:
\begin{equation}
\oint_\G G(z-w)\log(v(w))\dfrac{dw}{2i\pi}=-\sum_{n=1}^\infty\qf^nl_n(z-\e_1)=-\log(v(z-\e_1)).
\end{equation} 

A similar argument can be employed to treat $\hv(z)$. We decompose
\begin{equation}
\log(\hv(z))=\sum_{n=1}\qf^n\hl_n(z),
\end{equation} 
where the functions $\hl_n(z)$ have poles at $z=a_l-i\e_1$ for $i\in\mathbb{Z}^{>0}$, i.e. nowhere in the integration contour $\G$. Considering
\begin{equation}
\oint_\G G(z-w)\log(\hv(w))\dfrac{dw}{2i\pi}=\sum_{n=1}^\infty\qf^n\left[\oint_\G \dfrac{\hl_n(w)}{z-w+\e_1}\dfrac{dw}{2i\pi}-\oint_\G \dfrac{\hl_n(w)}{z-w-\e_1}\dfrac{dw}{2i\pi}\right],
\end{equation}
the first integral has only one pole inside $\G$ at $w=z+\e_1$, while the second integral has no pole at all. As a result,
\begin{equation}
\oint_\G G(z-w)\log(\hv(w))\dfrac{dw}{2i\pi}=-\sum_{n=1}^\infty\qf^n\hl_n(z+\e_1)=-\log(\hv(z+\e_1)).
\end{equation}

\bibliographystyle{utphys}

\providecommand{\href}[2]{#2}\begingroup\raggedright\endgroup

\end{document}